%Paper: hep-th/9503107
%From: Andreas Recknagel <anderl@itp.phys.ethz.ch>
%Date: Thu, 16 Mar 1995 15:05:40 +0100
%Date (revised): Tue, 23 May 1995 16:47:58 +0200

\magnification=\magstep1  \overfullrule=0pt
\advance\hoffset by -0.35truecm
\def\qed{{\vrule height4pt width4pt depth1pt}}
\def\lb{\lbrack}\def\rb{\rbrack}  \def\q#1{$\lb#1\rb$}
\def\pl{{\rm Phys.\ Lett.\ B\ }} 
\def\CMP{{\rm Commun.\ Math.\ Phys.\ }} \def\cmp{\CMP}

\def\bn{\bigskip\noindent} \def\mn{\medskip\smallskip\noindent}
\def\sn{\smallskip\noindent} 
\def\mod{{\rm mod}}  \def\Li{\hbox{\rm Li}}
 \def\eg{e.g.\ }  \def\lra{\longrightarrow}
\def\Z{\hbox{{\rm Z{\hbox to 3pt{\hss\rm Z}}}}}
\def\C{{\rm C \kern-5.5pt I \ }}
\def\un{\hbox{${\cal U}_N$}} 
\def\unp{\hbox{${\cal U}_{N'}$}}
\def\uinf{\hbox{${\cal U}_{\infty}$}}\def\calt{{\cal T}}
\def\uupinf{\hbox{${\cal U}^{\infty}$}}
\def\tn{\hbox{${\cal T}_N$}}
\def\for{\hbox{{\rm for}}} \def\csa{\hbox{$C^*$-algebra}}
\def\one{{\bf 1}} \def\N{^{(N)}}\def\upmo{^{-1}}
\font\gross=cmr10 scaled \magstep2
%%%%%%%%%%%%%%%%%%%%%%%%%%%%%%%%%%%%%%%%%%%%%%%%%%%%%%%%%
{\nopagenumbers
\rightline{ETH-TH/95-8}
\sn\rightline{hep-th/9503107}
\bn\bn\bn
\centerline{{\gross
The embedding structure and the shift operator}}
\sn
\centerline{{\gross of the U(1) lattice current algebra}}
\bn\bn\bn
\vfill \centerline{by}\bn\bn \vfill
\centerline{{ A.Yu.\ Alekseev}$\,{}^*$ and { A.\ Recknagel}}
\bn\bn
\centerline{Institut f\"ur Theoretische Physik, ETH-H\"onggerberg}
\centerline{CH-8093 Z\"urich, Switzerland}
\bn\bn\vfill\vfill\vfill\bn
\centerline{{\bf Abstract}}
\bn\bn
The structure of block-spin embeddings of the U(1) lattice current
algebra is described. For an odd number of lattice sites, the inner
realizations of the shift automorphism are classified. We present a
particular inner shift operator which admits a factorization involving
quantum dilogarithms analogous to the results of Faddeev and Volkov.
\bn\bn
\vfill\vfill\vfill
\sn
e-mail addresses: alekseev@itp.phys.ethz.ch, anderl@itp.phys.ethz.ch
\mn
${}^*$ On leave of absence from Steklov Mathematical Institute,
St.\ Petersburg \hfill
\eject}
%%%%%%%%%%%%%%%%%%%%%%%%%%%%%%%%%%%%%%%%%%%%%%%%%%%%%%%%%%%
\pageno=1
\leftline{\bf 1. Introduction}\bn
For $N\geq 3$, the U(1) lattice Kac Moody algebra is generated by the
operators \one\ and  $u_i$, $i = 1,\ldots,N$, with relations
$$\eqalign{&u_i u_j = u_j u_i\quad \for\ |i-j|\geq 2,  \cr
           &u_i u_{i+1} = q\,u_{i+1}u_i\quad\for\ i=1,\ldots,N \cr}
   \eqno(1)$$
with some $q\in\C$; $u_{N+i}\equiv u_i$.  We specialize to the case that
$q$ is a primitive $p\,$th root of unity, $p>1$. Then $u_i^* := u_i^{-1}$
defines a ${}^*$-operation; moreover, since the elements $u_i^p$
are all central, one can impose the further relations
$$ u_i^p = \one\ \ \hbox{{\rm for all}}\ i\,. \eqno(2)$$
Relations (1) and (2) define a finite-dimensional \csa\ \un\ of dimension
$p^N\,$. The centre $Z(\un)$ of \un\ is $p\,$-dimensional with generator
$$ c  = u_1u_2\cdots u_N    \eqno(3)$$
for odd $N$, but $p^2$-dimensional with generators
$$ c_1 = u_1u_3\cdots u_{N-1},\ c_2 = u_2u_4\cdots u_N \eqno(4)$$
for even $N$.
\sn
The U(1) lattice Kac Moody algebra, as the name indicates, is
designed as a
discretization of the commutation relations of a chiral abelian current.
Abelian and non-abelian lattice Kac Moody algebras have been
studied e.g.\ in \q{1-4}. They prove to be useful in investigating
the correspondence of the representation theory of quantum groups and of
Kac Moody algebras \q{3,4}. More recently, in \q{2} it was
shown that abelian lattice Kac Moody algebras are relevant
in the analysis of integrable field theories in two dimensions.
\sn
One of the main objects in the study of the above algebra is the
(periodic) shift automorphism
$$ s: \un\lra\un\,,\ u_i\longmapsto u_{i+1}\ , \eqno(5)$$
which corresponds to the chiral time evolution operator on the lattice,
see \q{1,2}. An important issue in those references was to find a
realization of $s$ by an inner automorphism: $s= ad_U$ for
some unitary $U\in\un$. Since inner automorphisms leave
central elements fixed, (3) and (4) make it clear that the
solution $U$ of this problem depends very much on whether $N$
is even or odd. Indeed, for even $N$ an inner shift operator
can only exist in a representation of \un\ where $c_1=c_2$, see
\q{1,2}. Whereas this condition may be considered in the framework
of field theory on the lattice with fixed (even) number of sites,
we are mainly interested in the continuum limit $N\to\infty$. The analysis
of algebra embeddings induced by refining the lattice (see Section 2)
shows that the structure of this continuum limit is in a sense much
closer to that of the algebras\un\ with $N$ odd than with $N$ even. In
particular, the constraint $c_1=c_2$ is incompatible with the embeddings.
So from this point of view, it appears to be important to construct
the shift operator for $N$ odd, which will be done in Section 3 of
this letter. In Section 2, we will make the above statements on the
continuum limit of the U(1) lattice Kac Moody algebra more precise.
\sn
Let us first introduce some notational convention: In many
equations below, factors $q^{1/2}$ will appear. For $p$ even,
this cannot be avoided since the spectrum of the central element $c$  is
contained in the $p\,$th roots of $-1$. The choice of the sign
of $q^{1/2}$ has no effect. For $p$ odd, we define $q^{1/2}$ with
the help of $2\upmo = {p+1\over2}$ in the ring $\Z_p\,$ so that in
this case everything could be written without any square roots
of $q$. We introduce them nevertheless because this enables us to
use identical formulas for $p$ even and odd in almost every equation.
\bn\bn
\leftline{\bf 2. Embeddings and the continuum limit algebra}
\bn
In \q{4}, it was shown how \un\ can be embedded into \unp\
for $N'>N$. The key idea is to associate
the generators $u_i$ of \un\ to the intervals $I_i \subset S^1$,
$i=1,\ldots,N$, of a triangulation ${\cal T}$ of the circle
into $|{\cal T}|:=N$ intervals. The procedure of refining ${\cal T}$
to $\calt'$ with $|\calt'|=N'>N$ by splitting the intervals, e.g.\
$I_i = I'_{i_1} \cup\ldots\cup I'_{i_{k_i}}$ with
$N'=\sum_{i=1}^N k_i$,  induces the following unital algebra
homomorphism
\def\io{\iota^N_{N'}}  \def\iott{\iota^{\calt}_{\calt'}}
\def\coll{\bigcup_{{\cal T}}\,{\cal U}_{\cal T}}
\def\ucalt{{\cal U}_{\cal T}}
$$\iott\,:\ {\cal U}_{\calt} \lra {\cal U}_{\calt'}\ ,\ \
  u_i^{\calt} \longmapsto q^{{k_i-1\over2}}u_{i_{k_i}}^{\calt'}
 \cdots u_{i_1}^{\calt'}\ .
    \eqno(6) $$
Here we have labelled the algebras by their corresponding triangulations,
and we have added a $q$-factor in comparison to \q{4} in order to
preserve relation (2).
\sn
Since the set of all triangulations of $S^1$ is directed (i.e.\
for any two $\calt$ and $\calt'$ there is a common refining
$\calt''\,$), the collection of all ${\cal U}_{\calt}$ and
all $\iott$ forms a directed system and one can define the
inductive limit
$$ \uupinf = \lim_{{\lra\atop{\cal T}}}\,{\cal U}_{\cal T}\ .\eqno(7)$$
$\uupinf$ may be understood as
$\bigl(\bigcup_{{\cal T}}\,{\cal U}_{\cal T}\bigr)\,/\sim\,$,
where $x_{\calt}\in{\cal U}_{\cal T}$ and $y_{\calt'}\in
{\cal U}_{\cal T'}$ are equivalent, $x_{\calt}\sim y_{\calt'}\,$,
iff there exists a common refining $\calt''$ of $\calt$ and
$\calt'$ such that $\iota^{\calt}_{\calt''}(x_{\calt})=
\iota^{\calt'}_{\calt''}(y_{\calt'})$ in ${\cal U}_{\cal T''}\,$.
\sn
As was explained in \q{4}, this algebraic version of the
continuum limit also carries a representation of
the chiral conformal group Diff${}_+(S^1)$, which is installed
more or less ``by brute force'' since the collection $\coll$
contains one algebra $\ucalt$ per triangulation, in particular
both $\ucalt$ and ${\cal U}_{g\calt}$ for any conformal
transformation $g\in {\rm Diff}_+(S^1)$.
\sn
At the same time, this implies that $\uupinf$ is rather difficult to handle,
simply because of its size: For a fixed length $N$ of triangulations,
there is already a continuum of associated algebras $\ucalt$
in the collection, which by definition are all isomorphic to \un.
Any two of these -- ${\cal U}_{\calt_1}$ and ${\cal U}_{\calt_2}$,
say, with $|\calt_1|=|\calt_2|=N$ -- are embedded into some
${\cal U}_{\calt'}$  -- one can take $|\calt'|=2N$ --, and standard
facts on finite-dimensional algebras tell that the images
$\iota^{\calt_1}_{\calt'}({\cal U}_{\calt_1})$ and
$\iota^{\calt_2}_{\calt'}({\cal U}_{\calt_2})$ are unitarily
equivalent in ${\cal U}_{\calt'}$ (see e.g.\ \q{5}), but they
need not coincide. Therefore we expect that also the inductive
limit $\uupinf$ is a huge object if the equivalence relation
$\sim$ from above is used.
\sn
In this paper, we will adopt a different notion of continuum
limit, which gives a smaller limiting algebra $\uinf$ and is
defined in closer analogy to what is done ``on the lattice''
when the number of sites is sent to infinity. We start from
a countable system $(\calt_N)_{N=3}^{\infty}$ of triangulations
where $N=|\calt_N|$ and $\calt_{N+1}$ is a refining of $\calt_N$;
again we associate an algebra \un\ to $\calt_N$. This means that
we have replaced all the unitarily equivalent algebras $\ucalt$
with $|\calt|=N$ from (7) by just one ``reference copy'' \un.
Together with the embeddings $\io$ as in (6) -- with an obvious
change of notation from $\calt$ to $N$ --, we get a directed
system $(\un,\io)$ so that the inductive limit
$$ \uinf = \lim_{{\lra\atop{N}}}\un   \eqno(8)$$
is well defined. $\uinf$ in principle depends on the specific
sequence of triangulations; we have suppressed this fact in
the above notation because at least the isomorphism class of $\uinf$
does not, as the following propositions will show.
\sn
We would like to mention that $\uinf$ still allows for the appealing
physical interpretation of the refining-embedding procedure (6) as an
inverse block-spin transformation \q{4}; the limiting algebras
$\uupinf$, $\uinf$ by definition display block-spin transformation
invariance, which is a lattice remnant of full conformal invariance.
\mn
For the case $q^p=1$, the continuum limit (8) is very
handy, since all the \un\  are multi matrix algebras; in other
words, $\uinf$ is a  so-called AF-algebra after taking
the \hbox{$C^*$-closure}, see \eg
\q{5} -- ``AF'' short for ``approximately finite-dimensional''.
Therefore $\uinf$ is determined up to isomorphism by the Bratteli diagram
(see \eg \q{5,6}) of the inclusions \hbox{$\io: \un\lra\unp$}, which in turn
is given as soon as we know the decomposition of \un\ into simple matrix
factors, and how these are embedded into those of \unp\ under $\io$ \q{7}.
\mn
{\bf Proposition 1}
$$\eqalign{&\un = \bigoplus_{i=0}^{p-1} \,{\cal U}^i_N \quad\  {\sl with}
   \quad {\cal U}^i_N \cong M_{p^{{N-1\over2}}}(\C)\quad{\sl for}\
{\sl all} \ i\ {\sl if}\ N\ {\sl is}\ {\sl odd}\ , \cr
  &\un = \bigoplus_{k,l=0}^{p-1}  {\cal U}^{k,l}_N \quad {\sl with}
 \quad {\cal U}^{k,l}_N \cong M_{p^{{N-2\over2}}}(\C)\quad{\sl for}\
{\sl all} \ k,l\ {\sl if}\ N\ {\sl is}\ {\sl even}\ . \cr}$$
\sn
Proof: The number of simple factors in \un\ is just given by the
dimension of the centre. Let  $e_i\N$ ($i=0,\ldots,p-1$;
$N$ odd) resp.\ $e_{kl}\N$ ($k,l=0,\ldots,p-1$; $N$ even) denote the
minimal central projections of \un, i.e.\ ${\cal U}^i_N=e_i\N\,\un\ $and
${\cal U}^{k,l}_N = e_{kl}\N\ \un$. It
is an easy exercise in root of unity calculations to show that
$$\eqalignno{
 &e_i\N = {1\over p}\,\sum_{n=0}^{p-1} q^{in}q^{-{N-2\over2}n}
                            \,c^n\ ,   &(9) \cr
  &e_{kl}\N = {1\over p^2} \sum_{n,m=0}^{p-1} q^{kn+lm}
                              \,c_1^n c_2^m\ ;  &(10) \cr}$$
of course, the numeration of the minimal central projections is
arbitrary, the one we have chosen above (involving the $N-2$ in
the second $q$-factor in (9)) will lead to simple formulas for the
embeddings $\io$ later on. In order to show that all the $e_i\N$,
resp.\ all the $e_{kl}\N$, have the same rank, we just use the
automorphisms $R_1:\ u_i \mapsto q^{\delta_{i1}}u_i$ and
$R_2:\ u_i \mapsto q^{\delta_{i2}}u_i$ of \un\ to permute the minimal
central projections:
$$\eqalign{ &R_1(e_i\N) = e_{i+1}\N\,, \cr
     &R_1(e_{kl}\N) = e_{k+1,l}\N\,,\ R_2(e_{kl}\N)=e_{k,l+1}\N\ ;\cr}$$
the subscripts are understood modulo $p$. \hfill \qed
\mn
\noindent As for the embeddings, it is sufficient to study ``elementary''
ones $\iota^N := \iota^N_{N+1}$ associated with cutting just one
interval of \tn\ into two pieces.
\mn
{\bf Proposition 2}
$$\eqalign{\iota^N({\cal U}^{k,l}_N)\
&\subset\ {\cal U}^{k+l\,({\rm mod}\,p)}_{N+1}
\,,\quad N\  {\sl even}\ ,     \cr
\iota^N({\cal U}^i_N)\  &= \bigoplus_{k+l\equiv i\,({\rm mod}\,p)}
 \!\!\!\!\!\! {\cal U}^{k,l}_{N+1}\,,\quad N\ {\sl odd}\ . \cr}$$
\sn
Proof: Let $\delta^{(p)}_{n,m}$ denote the Kronecker symbol defined
as $\delta^{(p)}_{n,m}=1$ if $n\equiv m\,({\rm mod}\,p)$ and zero
otherwise. For the special elementary embedding
$$ \eqalign{ &\iota^N(u_i^{(N)}) = u_i^{(N+1)}\ \ \for\
     i=1,\ldots,N-1\,, \cr
  &\iota^N(u_N^{(N)}) = q^{1/2}\, u^{(N+1)}_{N+1} u^{(N+1)}_N\,,  \cr}$$
one can verify by direct computation that
$$\eqalign{
\iota^N(e_{kl}^{(N)})\cdot e_i^{(N+1)}&=\delta^{(p)}_{k+l,i}\,
\iota^N(e_{kl}^{(N)})\,,\quad N\  {\rm even}\ ,     \cr
\iota^N(e_i^{(N)}) &= \sum_{k+l=i} e_{kl}^{(N+1)}\,,\quad N\
{\rm odd}\ .\cr}        \eqno(11)$$
Any other elementary embedding (cutting an interval other than $I_N\,$)
is obtained from this $\iota^N$ by applying shift automorphisms
$s^{(N)}, s^{(N+1)}$ as in (5) several times. But
$$ \eqalign{ s^{(N)}(e^{(N)}_i) = e^{(N)}_i\,\ \ N\ {\rm odd}\ , \cr
 s^{(N)}(e^{(N)}_{kl})=e^{(N)}_{lk}\,\ \ N\ {\rm even}\ ,\cr}\eqno(12)$$
so that the formulas (11)  are valid for arbitrary elementary
embeddings. By definition of the minimal central projections, the
equations (11)  imply the claim.   \hfill \qed
\mn
These two propositions show that the Bratteli diagram of $(\un,\io)$ in
(8) is independent of the chosen sequence of triangulations, and that it
splits into $p$ disconnected, identical subdiagrams, each of which looks
as follows: On the $N\,$th floor, $N$ odd, there is just one dot marked
$p^{{N-1\over2}}$ (representing one of the simple factors of \un, say
${\cal U}^i_N\,$); the $N+1\,$st floor consists of $p$ dots, each marked
$p^{{N+1-2\over2}}$ (which represent the factors ${\cal U}^{k,l}_{N+1}$
with $k+l\equiv i\,({\rm mod}\,p)\,$). From the single dot on the $N\,$th
floor, one line goes to every dot on the $N+1\,$st floor; from every such
dot there is one line to the single dot (marked $p^{{N+1\over2}}$)
on the $N+2\,$nd floor.
\sn
After ``telescoping'' the Bratteli diagram, see \q{5}, such that
only the odd floors with embeddings
$(\iota^{N+1}\circ\iota^N) :\ \un \lra
{\cal U}_{N+2}$ are left, the subdiagrams of \uinf\ become even simpler:
There is one dot per floor and $p$ lines between consecutive floors. One
can see this directly if one first shows that
$\iota^{N+1}(\iota^N(e_i^{(N)}))= e_i^{(N+2)}$
and then compares the rank
of the projections $e_i^{(N+2)}$ and $e_i^{(N)}$: The ratio equals  the
number of lines since $\iota^{N+1}\circ\iota^N$ is unital.
\sn
AF-algebras that possess an approximation by a tower of full matrix
algebras, as is the case for the subdiagrams of \uinf, are called
UHF-algebras (``uniformly hyperfinite''), see \q{5,8}. They are
completely determined (up to isomorphism) by giving the infinite
``product'' of numbers of lines in their Bratteli diagram;
in our particularly simple case we can write
$$ \uinf \cong \bigoplus_{\alpha=1}^p M_{p^{\infty}}\ .  \eqno(13)$$
Note that one can regard $M_{p^{\infty}}$ as an infinite tensor product
$M_{p^{\infty}}=M_p(\C)^{\otimes^{\infty}}$.
\sn
All in all, we have shown that it is indeed the algebras \un\ with
$N$ odd that determine the essential features of the continuum limit
as defined in (8); in particular, the centre of \uinf\ is spanned by
the images (in $\uinf\,$) of the $e_i\N$ and is also $p\,$-dimensional.
\bn\bn
\leftline{\bf 3. Construction of the shift operator}
\bn
In \q{1,2}, an expression for an inner shift operator $U\in\un$
with $s= ad_U$ was given for the case $N$ even under the additional
restriction $c_1=c_2$:
\def\even{{\rm even}} 
$$U_{\even} = r(u_2)r(u_3)\cdots r(u_{N})  \eqno(14)$$
The main feature of this solution is the appearance of the
function $r(w)$ which is determined up to a constant factor
by the functional equation ($\gamma\in\C$)
$$ w r(w)=  \gamma\, r(q^{-1}w)   \ ;\eqno(15)$$
here $q$ need not be a root of unity.
The function $r(w)$ enjoys the property $r(w)=r(w^{-1})$ and for
$\gamma = q^{1/2}$ can be factorized as
$$ r(w) = S(w)S(w^{-1})\quad{\rm with}\quad
   {S(q^{1/2}w)\over S(q^{-1/2}w)} = {1\over 1+w}\ .  \eqno(16)$$
If $q=e^{\epsilon}, |q|<1$, one can show that \q{9}
$$ S(w) = \exp\lbrace {1\over\epsilon}\,\Li_2(-w) + {\cal O}(\epsilon)
    \rbrace   $$
where
$$ \Li_2(w) = \sum_{k=1}^{\infty}{w^k\over k^2}      $$
is the Euler dilogarithm function; that is why $S(w)$ is called
``quantum dilogarithm''.
\mn
In the following, two constructions for a unitary shift operator of \un,
$N$ odd, will be given,  the first one starting from formula (14)
and producing a ``correction factor'', the second one
giving a formula for the most general unitary shift operator.
We would like to point out in advance that, although abstractly
$U$ is unique up to invertible central factors, the concrete
expression for it is highly non-unique. This makes the
existence of the beautiful formula (14) for even $N$ only the more
remarkable. The ones for $N$ odd turn out to be somewhat more complicated.
\mn
Besides the function $r(w)$ from (15) we will need $\tilde r(w)$
satisfying the similar relation
$$ w\tilde r(w)=\tilde\gamma\,\tilde r(q^{-2}w)    \eqno(17)$$
for some $\tilde\gamma\in\C$. Both functions may exist for
$q^p \neq 1$ as well (see \q{1,2} for explicit solutions),
and accordingly some of the formulas in Proposition 3 below are
valid whenever invertible solutions to (15) and (17) exist. But we
are mainly interested in the case when both $q^p=1$ and the
argument $w$ is such that $w^p = \one$. Then the constants
$\gamma$ and $\tilde\gamma$ cannot be chosen arbitrarily.
\sn
For $\gamma=q^{1/2}$ and $w^p = \one$, $w^*=w\upmo$,
$$ r(w) = {1\over\sqrt{p}}\sum_{k=0}^{p-1}q^{{k^2\over2}}w^k\eqno(18)$$
is a unitary solution of (15): $r(w)^*r(w)=\one\,$. The proof is
straightforward using the equation $\sum_{n=0}^{p-1} q^{kn}=
p\,\delta^{(p)}_{k,0}$ for a primitive root of unity.
\sn
The (up to constant factors) most general solutions of
(17) with $q^p=1$ and $w^p=\one$ are
$$\tilde r_l(w) = {1\over\sqrt{p}} \sum_{k=0}^{p-1}
  q^{{k(k+1)}}q^{-kl}w^k \ ,\quad \tilde\gamma_l = q^l  \eqno(19)$$
for $l\equiv 0,\ldots,p-1\ (\mod\,p)\, $, which again are
essentially theta functions.
With $w^*=w\upmo$, these $\tilde r_l(w)$ are unitary if
$p$ is odd, whereas we have
$$\tilde r_l(w)^* \tilde r_l(w)=\one +(-1)^{p/2+1-l}\,w^{p/2}\eqno(20)$$
for even $p$; note that then $q^{p/2}=-1$.
(20) shows that, for $p$ even, a unitary solution of (17) can
be found only for special arguments. For instance, assume that
$$ w^{p/2} = \sum_{i=0}^{p-1} (-1)^{i+1}e_i\N\quad \in Z(\un) \eqno(21)$$
where $e_i\N$ are the minimal central projections (9) of \un. Then
 $$ \tilde r(w):= {1\over\sqrt{2}} \sum_{i=0}^{p-1}
   \tilde r_{i+p/2}(e_i\N\,w)\eqno(22)$$
provides a unitary solution of (17). Here, in the different simple
 factors ${\cal U}_N^i$ of \un, different
$\tilde r_l(w)$ functions (and $\tilde\gamma_l$) have been used.
Now we are ready to state our first result on the shift operator:
\def\uonbf{{\bf u}_1} 
\mn
{\bf Proposition 3}\quad {\sl Denote by $W$ the product}
$$W=r(u_2)\,r(u_3)\cdots r(u_N)   \eqno(23)$$
{\sl and by $\uonbf$ the operator}
$$ \uonbf = \gamma\,u_Nu_{N-2}\cdots u_3\,u_1^{-1}u_2^{-1}u_4^{-1}
  \cdots u_{N-1}^{-1} \eqno(24)$$
{\sl with $\gamma=q^{1/2}$ as before.
A unitary inner shift operator $U\in\un$
for $N$ odd, $q^p=1$, can be written in the form
$$U=V\cdot W\ ; \eqno(25) $$
for $p$ odd, $V$ is given by
$$V= \tilde r (\tilde\gamma q^{-2}\uonbf) \eqno(26)$$
where $\tilde r(w)$ can be any $\tilde r_l(w)$
of the solutions  (19) of the relation (17)
with the associated factors $\tilde\gamma_l$ from
the functional equation;} \hfill \break
\noindent {\sl for $p$ even, $V$ is given by}
$$ V= \Bigl(\, \sum_{i=0}^{p-1}e_i\N\, x^{i+p/2}\Bigr)\,
  \tilde r(q^{-2}\uonbf)   \eqno(27)$$
{\sl with the function $\tilde r(w)$ defined in (22)  and}
$$x= (u_3u_5\cdots u_N)\upmo\ .  $$
{\sl If $q^p\neq1$ and invertible solutions of (15) and (17) exist,
the formula of the $p$ odd case can be taken to define an invertible
inner shift operator of the U(1) lattice Kac Moody algebra.}
\mn
Proof: From (1) and the functional equation (15) one deduces that
$$\eqalign{&r(u_i)u_{i-1}r(u_i)\upmo = \gamma u_iu_{i-1}\ ,\cr
 &r(u_i)u_{i+1}r(u_i)\upmo = \gamma u_i^{-1}u_{i+1}\ ,\cr
 &r(u_i)u_{j}r(u_i)\upmo = u_j\quad \for\ |\,i-j|\geq2\ .\cr}\eqno(28)$$
This in turn implies that the operator $W\in \un$ is almost the desired
shift, namely
$$\eqalign{&W u_i W^{-1} = u_{i+1}\quad \for\ i=2,\ldots,N-1\ ,\cr
   &W u_N W^{-1} = u_{1}\,\uonbf \  ,\cr
   &W u_1 W^{-1} = \uonbf^{-1}\,u_{2}\ ,\cr  }$$
with $\uonbf$ as in (24). $\uonbf$ commutes
with $u_i$ for $i=3,\ldots,N$,
but
$$\eqalign{ u_1 \uonbf &= q^{-2}\,\uonbf\, u_1\ , \cr
            u_2 \uonbf &= q^2\,\uonbf\, u_2\ . \cr}$$
Together with (17) this leads to relations similar
to (28):
$$\eqalign{ \tilde r(\uonbf)u_1\tilde r(\uonbf)\upmo &= u_1
      (\tilde\gamma\upmo q^2 \uonbf)^{-1}\ ,\cr
 \tilde r(\uonbf)u_2\tilde r(\uonbf)\upmo &=
  (\tilde\gamma\upmo q^2 \uonbf)u_2\ ,\cr}
    \eqno(29) $$
so that $U = \tilde r\bigl(\tilde\gamma q^{-2}\uonbf \bigr)\,W$
indeed satisfies the shift relation $Uu_i=u_{i+1}U$ for all $u_i$.
Since $\uonbf$ is unitary and $\uonbf^p=1$, the first factor,
i.e.\ $V$ from (26), is well defined and unitary for $p$ odd.
\sn
If $p$ is even, however, we have to take into account
the extra $w^{p/2}$ term in (20): Fortunately, $\uonbf^{p/2}$ is
just given by the rhs of (21) and therefore
a unitary $\tilde r(\uonbf)$ and also $\tilde r(q^{-2}\uonbf)$ can
 be defined
as in (22). Then the relations (29) have to be read in each simple
factor of \un\  separately, with $\tilde\gamma_{i+p/2}=q^{i+p/2}$
in the factor ${\cal U}_N^i$. This leaves us with an operator
$U':= \tilde r(q^{-2} \uonbf)W$ shifting $u_2,\ldots,u_{N-1}$
in the correct way
but producing extra ($i$-dependent) $\tilde\gamma$ factors when
acting on $u_N, u_1\,$:
$$\eqalign{&U'u_N {U'}\upmo\cdot e_i\N = \tilde\gamma_{i+p/2}\,u_1\cdot
     e_i\N\ , \cr
   &U'u_1 {U'}\upmo\cdot e_i\N = \tilde\gamma_{i+p/2}\upmo\,\,u_2\cdot
     e_i\N\ , \cr}$$
One finds that these $\tilde\gamma$ factors
cannot be absorbed by changing the argument of $\tilde r$ as
we did for $p$ odd, since this would spoil the property (21) and
would render $\tilde r(\tilde\gamma q^{-2}\uonbf)$ non-unitary.
Instead, we seek a unitary  operator $x\in\un$ with the properties
$$\eqalign{  &xu_ix\upmo = u_i\quad \for\ i=3,\ldots,N\,,\cr
 &xu_1x\upmo = q\upmo\, u_1 \,,\cr
  &xu_2x\upmo = q\, u_2 \,, \cr}$$
e.g.\ $x$ as in the Proposition. Note that $y:= u_2u_4\cdots u_{N-1}u_1$
satisfies the same commutation relations, and that $\uonbf\sim(yx)\upmo$,
$c\sim yx\upmo$.  After splitting into the simple factors, the
unitary operator
$$\sum_{i=0}^{p-1}e_i\N\, x^{i+p/2}$$
then acts by cancelling the  superfluous $\tilde\gamma\,$s.
Altogether, we arrive at
formula (27) for $V$ in the case when $p$ is even.  \hfill\qed
\mn
For further remarks on the object $\uonbf$, we refer to the concluding
section. But let us mention here that the splitting
of $U$ into separate shift operators for the simple factors, which
was necessary for $p$ even, is not so surprising if one identifies
the simple factors with inequivalent superselection sectors of the
model. Of course, every sector has its own chiral time evolution operator
so that from this point of view the surprise is rather that the
splitting is not necessary if $p$ is odd.
\bn
Since the centre of \un\ is $p$-dimensional for $N$ odd, there is
a $p\,$-torus of unitary shift operators, and Proposition 3 provides
just one example. In the following, we will give a formula for
arbitrary shift operators by determining a ``basis'' of unitary
shift operators in the simple factors ${\cal U}_N^i$; these shifts
are unique up to a phase.
\mn
We introduce some notation first: Let $d= {\rm gcd}(N,p)$ be the
greatest common divisor of $N$ and $p$, and $N=N'd$, $p=p'd$.
The natural $\Z_N$-action on $\Z_p$ induces a partition of
$\{0,\ldots,p-1\}$ into $d$ cycles $C_j=\{j,j+d,\ldots,j+(p'-1)d\}\,$,
$j=0,\ldots,d-1$; put differently, the $C_j$ are the cosets
$\Z_p/C_0\,$.
\def\modp{\ (\mod\,p)}
\mn
{\bf Proposition 4}\quad {\sl For $k=0,\dots,p-1\,$, let $U^{(k)}$ be
the operator}
$$\eqalignno{
 U^{(k)} &= {1\over\sqrt{d}}\,{1\over p^{{N-1\over2}}}\!
     \sum_{r_1,\ldots,r_N=0}^{p-1}\!\!\!\!\!\!\!\!\!{}^{\lb k\rb}\,\,
              q^{e(r_1,\ldots,r_N)}\, u_1^{r_1}\cdots u_N^{r_N}
                \quad \in \un      &(30) \cr
\noalign{\leftline{{\sl where }\hfill}}
 e(r_1,\ldots,r_N)&= -r_2(r_1+r_N)-r_3(r_2+r_1+r_N)-\ldots & \cr
  &\phantom{xxxxxxxxxxxx} - r_{N-1}(r_{N-2}+\ldots+r_1+r_N) &(31)\cr}$$
{\sl and $\sum \!{}^{\lb k\rb}$ indicates that the summation
variables are subject to $\sum_i r_i\equiv k\modp\,$.\hfill\break
\noindent If
$i\in C_j\,$, then $U^{(j/2)}e_i^{(N)}$ is the (up to a phase
unique) unitary shift operator of the simple factor ${\cal U}_N^i$
of \un\ for $i=0,\ldots,p-1\,$; we use $2\upmo ={d+1\over2}$ in $\Z_d$
to define $j/2$ if $j$ is odd.}   \hfill\break
\noindent{\sl If $U$ is an arbitrary unitary shift operator of \un,
there exist complex numbers $\lambda_i$ with $|\lambda_i|^2=1$ for
$i=0,\ldots,p-1$ such that}
$$U=\sum_{j=0}^{d-1} \Bigl( \sum_{i\in C_j} \lambda_i e_i^{(N)} \Bigr)
     U^{(j/2)}\ .     \eqno(32) $$
\mn
Remark: For definiteness, we have labelled the cycles $C_j$ by the
particular representative $j\in C_j$, but of course any other
would do as well. Accordingly, in the Proposition one can replace
$U^{(j/2)}$ by any $U^{(j'/2)}$ with $j'\in C_j$; note, however,
that the coefficients $\lambda_i$ in (32) depend on the basis of
shift operators in the simple factors.
\mn
Proof: We first have to show that the operators $ U^{(k)}$ from
(30,31) satisfy the relations
$$  U^{(k)}u_i=u_{i+1} U^{(k)} \eqno(33)$$
with all the generators $u_i$ of \un:
Suppose e.g.\ that $i=2,\ldots,N-1$; then
$$  u_1^{r_1}\cdots u_i^{r_i}u_{i+1}^{r_{i+1}}\cdots u_N^{r_N}\cdot u_i
 = q^{r_i'-r_{i+1}'-1}\, u_{i+1}\cdot
   u_1^{r_1}\cdots u_i^{r_i'}u_{i+1}^{r_{i+1}'}\cdots u_N^{r_N} $$
with $r_i' := r_i+1$, $r_{i+1}' := r_{i+1}-1$; this substitution
does not interfere with the summation restriction in
$\sum \!{}^{\lb k\rb}$, and furthermore
$$ e(r_1,\ldots,r_i,r_{i+1},\ldots,r_N) + r_i'-r_{i+1}'-1 =
   e(r_1,\ldots,r_i',r_{i+1}',\ldots,r_N)\ .$$
The cases $i=1,N$ are worked out similarly. Note that one way to arrive
at the operators $ U^{(k)}$ from the Proposition is just to make an
Ansatz for it with $e(r_1,\ldots,r_N)$ given by some undetermined
quadratic form, and then to solve for (33) -- leading to (31) as a
particularly simple solution.    \hfill\break
\noindent Thus the operators $U^{(k)}$ have the right shift commutation
relations but they are not unitary unless $d=1$: (33) implies that
$ U^{(k)*} U^{(k)} \in Z(\un)$ so that we have to determine the
real numbers $\mu^{(k)}_i$ in the decomposition
$$ U^{(k)*} U^{(k)} = \sum_{i=0}^{p-1}\mu^{(k)}_i e_i\N  \eqno(34)$$
in order to see which $U^{(k)}$ projects onto a unitary shift operator in
which simple factor ${\cal U}_N^i$. The computation of $\mu^{(k)}_i$
is a little tedious, so we will only list the various steps:
One first inserts (30,31) into the lhs of (34), uses the summation
restrictions to eliminate two of the $2N$ summation variables, and
brings the product of $u_i\,$s into the standard form as in (30) again;
here additional $q$-factors arise. Next a substitution of
summation variables is applied such that only half of the new ones
appear in the exponents of \un-generators.
It turns out that the other $N-1$  summations can be performed
with the help of $\sum_{k=0}^{p-1}q^{kn} =p\,\delta^{(p)}_{n,0}$,
which on the whole yields a factor $p^{N-1}$ as well as conditions
on the remaining summation variables: Namely, they all have to be
equal (modulo $p$) to each other, and the last free variable, $r$ say,
is subject to $Nr\equiv 0 \modp$; this means that the sum is over
$r=0,p',\ldots,(d-1)p'$ only; we arrive at
\def\pdivr{{{r=0}\atop{p'|r}}}
$$U^{(k)*} U^{(k)} = {1\over d}  \sum_{\pdivr}^{p-1}
   q^{e(r,\ldots,r)+2kr}\, u_1^r\cdots u_N^r \ . $$
With our definition (9) of the minimal central projections $e_i\N$, we have
$$ u_1^r\cdots u_N^r = \sum_{i=o}^{p-1} q^{{N-2\over2}r^2-ir}\,e_i\N $$
so that
$$\mu^{(k)}_i={1\over d}\sum_{\pdivr}^{p-1} q^{(2k-i)r}\ ;\eqno(35)$$
here we have used $Nr\equiv0\modp$ to simplify the $q$-exponent. This
condition, by the way, also shows that $\mu^{(k)}_{i+N \modp} =
\mu^{(k)}_i$, which means that $\mu^{(k)}_i$ depends only on
the cycle $C_i$ and not on the representative
$i\in C_i$.  Finally, note that the sum in (35) ranges over all $d\,$th
roots of unity, raised to the power $2k-i$, therefore
$$ \mu^{(k)}_i = \delta^{(d)}_{2k,i} \ ;$$
in particular, $\mu^{(k)}_i =1$ for $k=i/2$ as claimed. \hfill\qed
\bn
It would be desirable to rewrite the formulas for the shift
operators given above into a product of $r$-functions similar
to Proposition 3, if possible with more symmetric arguments. This
would allow us to make contact with the quantum dilogarithm and in
addition would perhaps lead to expressions valid for $q^p\neq1$ as well.
Unfortunately, we have not been able to find a nice factorization
of a shift operator for general $N$ yet.  The difficulties start with
choosing an appropriate unitary shift operator from the $p\,$-torus
of solutions given above. Then, it is advisable to approach
the factorization problem from a general point of view -- since by simply
``guessing'' arguments of $r$-functions which are to be split off
an expression like (30) one typically arrives at non-factorizable
``remainders''. Generally, factorization of  an operator like the
$U^{(k)}$ in Proposition 4 is achieved if we can find new summation
variables which diagonalize the quadratic form in the $q$-exponent.
However, the latter is then not just given by $e(r_1,\ldots,r_N)$ from
(31), but there are additional $q$-factors arising from re-grouping
the generators $u_i$ into more complicated expressions, which are
in turn dictated by the substitution of variables.
All in all, this ``$q$-diagonalization problem'' leads to equations
considerably more involved than in the commutative case.
\bn\bn   \leftline{\bf 4. Further comments}   \bn
The main part of this paper was addressed to the construction of
the shift operator of \un\  for $N$ odd -- not merely in order to
complete the picture begun in \q{1,2}, but rather because the
considerations of Section 2 suggest that $N$ odd is the
more important case. Unfortunately, the results we have obtained do
not look as nice as formula (14) for the $N$ even case -- where,
however, some complications are hidden in the necessary condition
$c_1=c_2$. Indeed, the only difference between formulas (14) and
(25) is the additional factor involving the operator $\uonbf$ in
the latter; but observe that $\uonbf \sim \iota^{N-1}(c_2c_1\upmo)\,$
where $\iota^{N-1}$ is an elementary
embedding $\iota^{N-1}:{\cal U}_{N-1}\lra \un$ induced by cutting
the interval $I_1$ of $\calt_{N-1}$ into two pieces, compare (6).
Thus, $\uonbf$ appears simply because $c_1\neq c_2$ in general. Stated
differently, if we were to restrict $U$ as in (25) to the subalgebra
$\iota^{N-1}({\cal U}_{N-1})\subset\un$\ and to impose $c_1=c_2$
in ${\cal U}_{N-1}\,$, the ``correction factor'' $V$ from Proposition 3
would disappear. This fact is also evidence for the consistency of
the shift operator construction with the embedding structure.
\sn
If one is interested in the continuum limit of the model and of
the shift operator -- a continuum limit defined by the embedding
procedure of \q{4} --, then one is forced to consider the
shift operator for $N$ odd since the condition $c_1^{(N-1)}=c_2^{(N-1)}$
is not mapped to $c_1^{(N+1)}=c_2^{(N+1)}$ by $\iota^N\circ\iota^{N-1}\,$.
One may expect that the $N$ odd expression (25) will approach
a simpler limit when $N\to\infty$, because we would like to think
of $\uonbf$ as some kind of ``lattice artefact'': to $u_1$ a string
of $u_iu_{i+1}\upmo$ is attached which hopefully ``averages
out'' in the limit of large $N$.
\sn We think that also the formulas given in Proposition 4 might prove
useful in this context, e.g.\ if one wants to check rigorously
whether the shift operators of \un\ have some limit in $\uinf$
(or affiliated to this algebra).
\mn
Investigating this continuum limit further is in itself an interesting
project. Among the obvious problems are to clarify the relation
between $\uupinf$ of \q{4} and the smaller algebra $\uinf$ used here,
and to decide whether the full conformal symmetry is restored in
$\uinf$. For field theoretic applications it is very important to
find representations with proper ground states.
\mn
One might also be able to construct a direct link
between the (limiting) shift operator and the $L_0$-component of
the energy momentum tensor. Since the shift operator can be
expressed through the quantum dilogarithm (16), such a connection
could e.g.\  be useful to find generalizations of the existing
(classical) dilogarithm identities for the central charge and the
conformal dimensions of some conformal field theories, see e.g.\
references \q{10}. On the other hand, there are  quantum dilogarithm
formulas for the scattering matrices of some 1+1-dimensional
integrable quantum field theories \q{1,2,11}. Because of all these
connections, it appears to be important to explore this function further.
\mn
Let us finally remark that the
system of algebras \un\  may also serve as a ``toy model'' for
applying the algebraic approach to quantum field theory, see e.g.\
\q{12} and references therein; in particular, one can
define ``local algebras'' associated
to proper intervals of $S^1$. We expect that, as soon as questions
concerning the continuum limit and its relevant representations
have been clarified, the U(1) lattice Kac Moody algebra could become
much more than a toy model and serve as a basis for
free field constructions to obtain other (conformal and maybe even
massive) field theories -- see \q{4} for further details. In this
context, the fact that our basic model leads to an equally basic operator
algebra is rather encouraging, since in the theory of operator
algebras there is a number of constructions leading from UHF-algebras
to more involved and more interesting algebras.
\bn \vfill \bn
We would like to thank  A.\ Fring, J.\ Fr\"ohlich,
M.\ R\"osgen, and in particular K.\ Gawedzki
for discussions and comments. We thank A.Yu.\ Volkov
who, after the main part of this work was completed, informed
us that he
has found a formula for inner shift operators for arbitrary $N$
by a different method \q{13}. \sn
A.R.\ is supported by a European Union ``Human
Capital and Mobility'' fellowship.
\bn
\eject
\parindent=20pt \def\q#1{\item{\rm #1.\hfill}}
\bn\bn\leftline{{\bf References }} \bn
\q{1}L.D.\ Faddeev, A.Yu.\ Volkov,
    {\it Abelian current algebra and the Virasoro algebra}
    {\it on the lattice},  \pl {\bf 315} (1993) 311 \sn
\q{2}L.D.\ Faddeev,
   {\it Current-like variables in massive and massless
   integrable models},
   Lectures given at Enrico Fermi School on Quantum Groups and Their
   Physical Applications, Varenna, Italy, 1994,  hep-th/9408041 \sn
\q{3}A.Yu.\ Alekseev, L.D.\ Faddeev, M.\ Semenov-Tian-Shansky,
   {\it Hidden quantum groups}
   {\it inside Kac-Moody algebras},  \cmp {\bf 149} (1992) 335 \sn
\q{4}K.\ Gawedzki,
   {\it Quantum group symmetries in conformal field theory}, IHES
   preprint 1992, hep-th/9210100; {\it CFT on the lattice},
   talk given at the DPG-Tagung, Mainz, Germany, 1993; \hfill
\item{}F.\ Falceto, K.\ Gawedzki, {\it Lattice Wess-Zumino-Witten
   model and quantum  groups},
   \hbox{J.\ Geom.\ Phys.\ {\bf 11} (1993) 251} \sn
\q{5}B.\ Blackadar, {\it K-Theory for Operator Algebras}, MSRI
   Publ.\ 5, Springer, New York 1986;  \hfill
\item{}E.G.\ Effros, {\it Dimensions and $C^*$-algebras},
   Reg.\ Conf.\ Ser.\ in Math.\ {\bf46}, Amer.\ Math.\ Soc.\ 1980 \sn
\q{6}F.M.\ Goodman, P.\ de la Harpe, V.F.R.\ Jones, {\it Coxeter
   Graphs and Towers}
   {\it of Algebras}, MSRI Publ.\ 14, Springer, New York 1989 \sn
\q{7}A.\ Recknagel, {\it AF-algebras in conformal field theory},
   talk given at the Satellite Colloquium on New Problems
   in the General Theory of Fields and Particles, Paris, 1994 \sn
\q{8}J.\ Glimm, {\it On a certain class of operator algebras},
   Trans.\ Amer.\ Math.\ Soc.\ {\bf95} (1960) 318  \sn
\q{9}L.D.\ Faddeev, R.M.\ Kashaev,
   {\it Quantum dilogarithm}, Mod.Phys.Lett.\ A {\bf9} (1994) 427 \sn
\q{10}A.\ Kuniba, T.\ Nakanishi, {\it Spectra in conformal field
   theories from the Rogers dilogarithm}, Mod.\ Phys.\ Lett.\ A {\bf7}
   (1992) 3487;      \hfill
\item{}W.\ Nahm, A.\ Recknagel, M.\ Terhoeven, {\it Dilogarithm
   identities in conformal field theory}, Mod.\ Phys.\ Lett.\ A {\bf8}
   (1993) 1835; \hfill
\item{}R.\ Kedem, T.R.\ Klassen, B.M.\ McCoy, E.\ Melzer,
   {\it Fermionic sum representations for conformal field theory
   characters}, Phys.\ Lett.\ B {\bf307}  (1993) 68    \sn
\q{11}A.\ Fring, private communication\sn
\q{12}R.\ Haag, {\it Local Quantum Physics}, Springer,
   Berlin-Heidelberg 1992 \sn
\q{13}A.Yu.\ Volkov, unpublished \sn
\bye